\documentclass[prb,twocolumn,showpacs]{revtex4}%
\usepackage{amsfonts}
\usepackage{amsmath}
\usepackage{amssymb}
\usepackage{graphicx}
\usepackage{bm}%
\providecommand{\U}[1]{\protect\rule{.1in}{.1in}}
\begin{document}
\title{Effect of field dependent core size on reversible magnetization of
high-$\kappa$ superconductors}
\author{V.~G. Kogan}
\affiliation{Ames Laboratory - DOE and Department of Physics and Astronomy, Iowa State
University, Ames IA 50011, USA}
\author{R. Prozorov}
\email[corresponding author: ]{prozorov@ameslab.gov}
\affiliation{Ames Laboratory - DOE and Department of Physics and
Astronomy, Iowa State University, Ames IA 50011, USA}
\author{S.~L. Bud'ko}
\affiliation{Ames Laboratory - DOE and Department of Physics and Astronomy, Iowa State
University, Ames IA 50011, USA}
\author{P.~C. Canfield,}
\affiliation{Ames Laboratory - DOE and Department of Physics and Astronomy, Iowa State
University, Ames IA 50011, USA}
\author{J.~R. Thompson,}
\affiliation{Materials Science and Technology Division, ORNL, Oak Ridge, TN 37831,}
\affiliation{and Department of Physics, University of Tennessee, Knoxville, Tennessee
3796-1200, USA}
\author{J. Karpinski and N.D. Zhigadlo,}
\affiliation{Solid State Physics Laboratory, ETH, 8093 Zurich, Switzerland}
\author{P. Miranovi\'{c}}
\affiliation{Department of physics, University of Montenegro, 81000 Podgorica, Serbia and Montenegro}
\date{\today}


\begin{abstract}
The field dependence of the vortex core size $\xi(B)$ is incorporated in the
London model, in order to describe reversible magnetization $M(B,T)$ for a
number of materials with large Ginzburg-Landau parameter $\kappa$. The
dependence $\xi(B)$ is directly related to deviations in $M(\ln B)$ from
linear behavior prescribed by the standard London model. A simple method to
extract $\xi(B)$ from the magnetization data is proposed. For most materials
examined, $\xi(B)$ so obtained decreases with increasing field and is in
qualitative agreement both with behavior extracted from $\mu$SR and small
angle neutron scattering data and with that predicted theoretically.

\end{abstract}
\maketitle

\section{Introduction}

\subsection{On the London model}

Despite its simplicity, the London approach is a powerful tool in describing
magnetic properties of the mixed state. In fact, short of the full-blown
microscopic theory, it is the only method available for low temperatures. The
approach is based on the London equation:
\begin{equation}
h-\lambda^{2}\nabla^{2}h=\phi_{0}\sum_{n}\delta({\bm r}-{\bm r}_{n})\,,
\label{London}%
\end{equation}
where $h(\bm r)$ is the magnetic field, $\lambda$ is the penetration depth (a
temperature dependent \textit{constant} in uniform samples), $\phi_{0}$ is the
flux quantum, and $\bm r_{n}$ are vortex positions. For simplicity, the
equation is written for isotropic materials. This approach fails at distances
of the order of the coherence length $\xi$; still, in materials with
$\kappa=\lambda/\xi\gg1$, there is a broad domain of intermediate fields
$\phi_{0}/\lambda^{2}\ll H\ll\phi_{0}/\xi^{2}$ where the complexity of the
vortex core contributions to the total energy can be disregarded and the
London approach suffices for the description of macroscopic magnetic properties.

As far as the equilibrium properties of the flux-line lattice are concerned,
the pivotal point is the expression for the free energy
\begin{equation}
{\tilde{F}}=F-\frac{B^{2}}{8\pi}=\frac{\phi_{0}B}{32\pi^{2}\lambda^{2}}%
\,\ln\frac{e\eta H_{c2}}{B}\,. \label{F_Lond}%
\end{equation}
The right-hand side here is the interaction energy of vortices forming a
periodic lattice, $B$ is the magnetic induction. This expression is obtained
by transforming the sum of pair-wise interactions of vortices to a sum over
the reciprocal lattice, see e.g., Ref.\thinspace\onlinecite{deGennes}. The sum
(or the integral over the reciprocal plane $\bm k$) is logarithmically
divergent so that a cutoff at $k\sim1/\xi\sim1/\rho_{c}$ is needed ($\rho_{c}$
is the size of the vortex core). This yields $\ln(\phi_{0}/2\pi\xi^{2}B)$. The
parameter $\eta$ of the order unity is commonly introduced to account for
uncertainty of the cutoff (along with the uncertainty in the lower limit of
the integral of the order of inverse intervortex spacing $\sim\sqrt{B/\phi
_{0}}$); $e=2.718$... appears in Eq.(\ref{F_Lond}) for convenience of not
having it in the expression for the magnetization. Again, the energy in the
form of Eq.(\ref{F_Lond}) holds in intermediate fields $H_{c1}\ll H\ll H_{c2}%
$, the domain existing only in materials with large Ginzburg-Landau parameter
$\kappa$ ($H_{c1}$ and $H_{c2}$ are the lower and upper critical fields).
Hence, although the length $\xi$ (or the core size $\rho_{c}$) does not appear
in the London equation (\ref{London}), it enters the energy expression
(\ref{F_Lond}) through the cutoff and therefore affects, presumably weakly,
macroscopic quantities such as the magnetization and other properties of the
mixed state.

Significant effort has recently been put in studies of the vortex core size
$\rho_{c}$; see the review by Sonier and references therein.\cite{Sonier1}
Notably, whatever definition of $\rho_{c}$ is adopted, the low temperature
$\rho_{c}$ (extracted from the $\mu$SR data) decreases with increasing applied
magnetic field in a number of materials such as NbSe$_{2}$, V$_{3}$Si,
LuNi$_{2}$B$_{2}$C, YBa$_{2}$Cu$_{3}$O$_{7-\delta}$, and CeRu$_{2}$; their
physical characteristics have little to do with one other, except that all of
them have a large GL parameter $\kappa=\lambda/\xi$ and exhibit large regions
of reversible magnetic behavior. One can add to this list a heavy fermion
compound CeCoIn$_{5}$, for which the interpretation of small angle neutron
scattering data (SANS) requires a similar behavior of the coherence
length.\cite{Morten} The dependencies $\rho_{c}(B)$ for all tested materials
are qualitatively similar: when the field increases toward $H_{c2}$, $\rho
_{c}(B)$ decreases roughly as $1/\sqrt{B}$. In other words, in large fields
$\rho_{c}$ is roughly proportional to the intervortex spacing.

A few qualitative reasons for the core shrinking with increasing field have
been discussed in literature; see the review \onlinecite{Sonier1}. Perhaps the
simplest defines the core boundary as a position where the divergent London
current $c\phi_{0}/8\pi^{2}\lambda^{2}r$ of an isolated vortex reaches the
depairing value, i.e., as $r\sim\xi$. In the mixed state, neighboring vortices
suppress the circulating current by contributing currents of the opposite
direction. Hence, the depairing value is reached at a shorter distance from
the vortex center, and consequently $\rho_{c}$ should decrease with increasing field.

The vortex core size, $\rho_{c}$, is of the order of the coherence length
$\xi$ and, in fact, it is often identified with $\xi(T)$. Strictly speaking,
the latter is defined only at the upper critical field: $H_{c2}\left(
T\right)  =\phi_{0}/2\pi\xi^{2}$. Nevertheless the length $\xi$ is used to
describe the mixed state at fields not necessarily close to $H_{c2}$. The
question of possible field dependence of $\xi$ has been considered by one of
us for the isotropic case.\cite{K85} It was shown that in the dirty limit one
can use $\xi=(\phi_{0}/2\pi H_{c2})^{1/2}$ at any field within the mixed
phase; the same is true near the critical temperature $T_{c}$ for any
scattering strength. However, in general, when the field is reduced below
$H_{c2}$, the value of $\xi(B)$ increases, an effect that is profound in clean
materials at low temperatures. Calculations of Ref.\thinspace\onlinecite{KZ2}
are in accord with the $\mu$SR results cited above. In the following, we
denote as $\xi_{c2}$ the value of $\xi(B)$ at $B=H_{c2}$ to stress that in
general $\xi(B)\neq\xi_{c2}$ for $B<H_{c2}$.

Another common theoretical definition of $\rho_{c}$ is based on the slope of
the order parameter $\Delta(r)$ at the vortex axis $r=0$, normalized either to
its value $\Delta(\infty)$ far from the single vortex or to the value of
$\Delta(a/2)$ half-way to the nearest neighbor in the mixed state: $1/\rho
_{c}=\Delta^{\prime}(0)/\Delta(a/2)$. Recent microscopic calculations of this
quantity by Miranovi\'{c} \textit{et al.} showed a variety of field dependent
behaviors of $\rho_{c}$ at low temperatures depending on the scattering
strength.\cite{Pedja} In particular, this work suggests that $\rho_{c}(B)$ may
have a minimum which, however, has not been seen in $\mu$SR experiments. There
are many different ways to define $\rho_{c}$ customized for different
experimental or theoretical needs (see, e.g., the discussion in Ref.\thinspace
\onlinecite{KZ2}). For the purpose of this paper, these differences are
irrelevant and we use the terms $\rho_{c}(T,B)$ and $\xi(T,B)$ as the same.

In the following we provide experimental data for the field dependence of the
reversible magnetization for a single crystal YNi$_{2}$B$_{2}$C in a broad
temperature region to demonstrate the known fact: the data cannot be described
by the standard London model. We then derive a closed form expression for the
cutoff $\xi(B)$ needed to represent correctly the data $M(B)$ with the help of
the London model. We show that $\xi(B)$ so chosen is qualitatively consistent
with the field dependence of $\xi$ recorded by $\mu$SR and discussed
theoretically. Finally, we demonstrate that the new model generates a
consistent description of the magnetization data for a number of unrelated
materials with large $\kappa$. The goal of this paper is to demonstrate that
$\xi(B)$ can be in principle extracted from the magnetization data, a less
demanding experimental procedure as compared to $\mu$SR or SANS. Within our
approach, \textit{the London penetration depth is field independent}, whereas
the field dependence of $\xi$ alone suffices to explain the data.

\section{Experimental Aspects}

\subsection{Sample preparation}

Single crystals of YNi$_{2}$B$_{2}$C, LuNi$_{2}$B$_{2}$C, and Lu(Ni$_{1-x}%
$Co$_{x})_{2}$B$_{2}$C were grown out of Ni$_{2}$B or (Ni$_{1-x}$Co$_{x})_{2}%
$B flux in a manner similar to the growth of other borocarbide crystals. As
discussed in Ref.\thinspace\onlinecite{Cheon}, Co doping serves as a
convenient tool to move from clean to dirty limit.

NbSe$_{2}$ crystals were grown via iodine vapor transport technique (REF) and
had $T_{c}\approx7.1$ K and the residual resistivity ratio $RRR\approx40$.

MgB$_{2}$ single crystals of submillimeter sizes were grown by a high pressure
technique as described in Ref.\thinspace\onlinecite{Karpinski}.

A single crystal of V$_{3}$Si was grown using a Bridgeman method, in which a
floating zone was created by rf induction heating. The samples for
investigation were cut by a wire saw and oriented using Laue method. Samples
had typical dimension of $2\times2\times4$ mm.

\subsection{Magnetic measurements}

The magnetization measurements were performed by using several Quantum Design
MPMS systems. In a typical experiment, a full $M\left(  H\right)  $ loop was
recorded and only its reversible part, above the irreversibility field,
$H_{irr}$, was used for the analysis. $H_{irr}$ was determined as a field
where ascending and descending branches coincided or were sufficiently close
(a weak hysteresis).%

\begin{figure}
[ptbh]
\begin{center}
\includegraphics[
height=7.0731cm,
width=9.1028cm
]%
{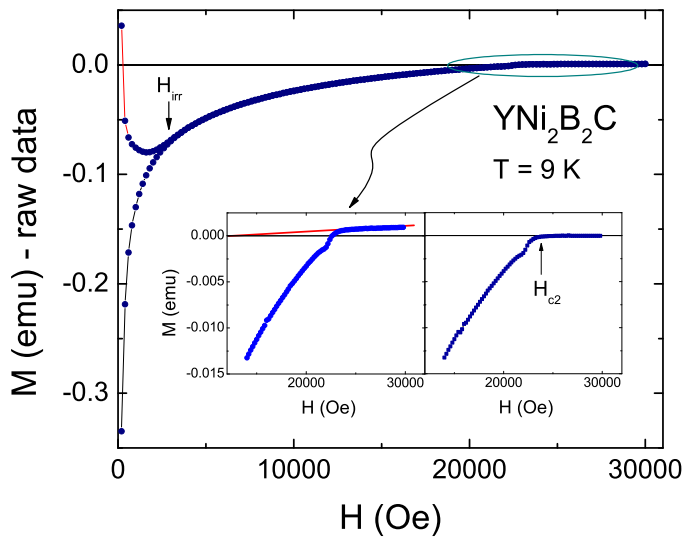}%
\caption{(color online) An example of $M(H)$ for YNi$_{2}$B$_{2}$C at
$T=9$\thinspace K. The main plot shows both up- and down-field scans and the
irreversibility field. The upper inset illustrtes how the normal state
paramagnetic contribution is subtracted with the result shown in the lower
inset. The upper critical fields $H_{c2}$ is indicated by an arrow.}%
\label{fRAW}%
\end{center}
\end{figure}

The procedure is demonstrated in Fig.\,\ref{fRAW}. The main frame shows raw
data with clear range of reversible behavior above $H_{irr}$ indicated by an
arrow. The left inset shows the expanded portion of the raw data in the
vicinity of $H_{c2}$. A small paramagnetic background (from the sample, the
sample holder and perhaps residual flux on the crystal surface) is clearly
seen as a linear-in-$H$ contribution. After this contribution is subtracted,
we obtain the superconducting diamagnetic signal shown in the right inset. The
$H_{c2}$ is indicated by an arrow. Superconducting transition temperature was
measured in a small ($H=10$\thinspace Oe) applied field. In large fields of
our interest, demagnetization effects are weak; in the following we do not
distinguish between the applied field $H$ and the induction $B$. As explained
below, we do not need the sample volume in our analysis so that we can use on
equal footing field dependencies of the magnetic moment (in $emu$) or of the
magnetization (in $G$); we use the notation $M$ for both quantities.

\section{Modified London Model}

The standard London energy (\ref{F_Lond}) gives an equilibrium magnetization
that is linear in $\ln B$ in intermediate fields:
\begin{equation}
M=-\frac{\partial\tilde{F}}{\partial B}=\frac{\phi_{0}}{32\pi^{2}\lambda^{2}%
}\,\ln\frac{\eta{\ H_{c2}}}{B}\,. \label{M_Lond}%
\end{equation}
Hence, the standard London model requires a plot of $M$ versus $\ln B$ to be a
straight line.

\subsection{YNi$_{2}$B$_{2}$C}

\begin{figure}
[tbh]
\begin{center}
\includegraphics[
height=7.3148cm,
width=9.1028cm
]%
{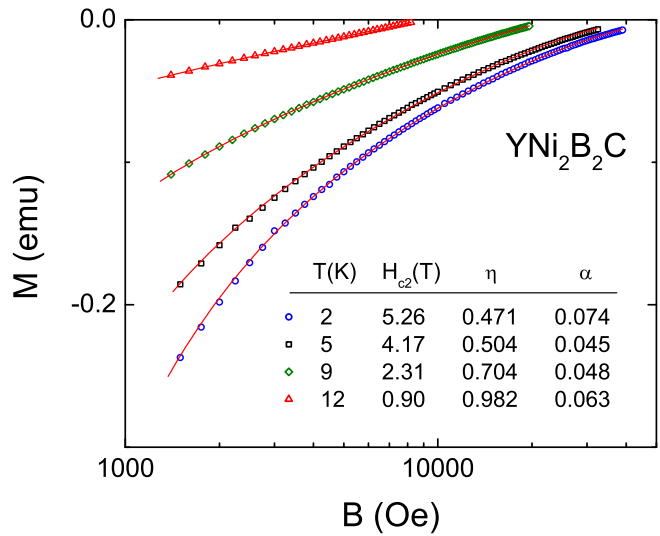}%
\caption{(color online) The magnetization $M(\ln B)$ for YNi$_{2}$B$_{2}$C at
$T=2$, 5, 9, and 12\thinspace K. The upper critical fields $H_{c2}$ are the
positions of kinks in $M(\ln B)$ not shown in the figure. The solid curves are
obtained by fitting the data to Eq.\thinspace(\ref{M_exp}) with the fitting
parameters $M_{0}$ (shown in the inset), $\eta$, and $\alpha$.}%
\label{fYNBC}%
\end{center}
\end{figure}

Fig.\,\ref{fYNBC} shows \textit{reversible} magnetization for a single crystal
YNi$_{2}$B$_{2}$C in fields parallel to the $c$ axis at $2,5,9$, and $12$\,K.
Clearly, the deviations from the London prediction increase with decreasing
$T$; at low temperatures $M(\ln B)$ is far from being linear.%

We note that in many materials with $\kappa\gg1$, it is difficult to
distinguish between a narrow Abrikosov domain near $H_{c2}$ with
$M\propto(H_{c2}-B)$ and a broad London domain where the magnetization depends
on the field in a slow, nearly logarithmic manner. For this reason the
Abrikosov part of $M(B)$ is sometimes discarded altogether; this amounts to
setting $\eta=1$ in Eq.\thinspace(\ref{M_Lond}).\cite{Tinkham} Of course, this
cannot be done for materials with $\kappa\sim1$. We will follow this
simplification in our analysis and indicate the cases when this cannot be done.

To formally account for deviations of the data from the behavior prescribed by
the standard London formula, we add to expression (\ref{M_Lond}) two
additional terms: const$/B$ (to correct for the low field behavior) and
const$\cdot B$ (to account for the high field curvature). These are, perhaps,
the simplest possible modifications one can think about.\cite{remark*} A
restriction upon the constants is provided by a requirement that $M(H_{c2}%
)=0$. Hence
\begin{align}
M= -M_{0}\left[  \ln\frac{ \eta H_{c2} }{ B }+\alpha\frac{ H_{c2} }{ B }%
-(\ln\eta+\alpha)\frac{B}{ H_{c2}}\right]  ; \label{M_exp}%
\end{align}
where $M_{0}=\phi_{0}/32\pi^{2}\lambda^{2} $. Certainly, the form
(\ref{M_exp}) is not the only possiblity for representing the available data.
Other forms were suggested in the literature\cite{Brandt,Hao,Kosh,nonloc}
based on different theories and assumptions. We stress that the expression
(\ref{M_exp}) is just an empirical formula to represent the data. We choose it
because of its simplicity, and because - as is demonstrated below - it is
sufficiently flexible to represent the magnetization data in a host of
materials with very different physical properties.

The solid lines in Fig.\thinspace\ref{fYNBC} are the data fits to
Eq.\thinspace(\ref{M_exp}). The value of the upper critical field for each $T$
is read directly from the raw data as explained in Fig.\thinspace\ref{fRAW}.
We are left with three fit parameters $M_{0},\eta$, and $\alpha$. Two of these
are shown in the table of Fig.\thinspace\ref{fYNBC}. The inset in
Fig.\thinspace\ref{fYNBCxi} shows that the $T$-dependence of $M_{0}%
\propto1/\lambda^{2}$ is qualitatively consistent with the behavior of the
superfluid density $\propto\lambda^{-2}$. The quality of the fits is good;
hence, the empirical form (\ref{M_exp}) can be used to represent the data
reasonably accurately.

Examining possible modifications of the London model to account for the
deviations of $M(\ln B)$ of Fig.\thinspace(\ref{fYNBC}) from linear behavior,
one should bear in mind the difference between the roles of two fundamental
lengths, $\lambda$ and $\xi$, within the London theory. The length $\lambda$
enters Eq.\thinspace(\ref{London}) which is the basis of the whole approach.
On the other hand, the length $\xi$ is absent in the London equation and
enters the energy expression as an uncertain cutoff used to mend the inherent
shortcoming of the London model. Therefore, considering possible modifications
of this model, one still has some freedom - however limited - in working with
$\xi$, unlike the case of $\lambda$.

Comparing the data of Fig\,(\ref{fYNBC}) with predictions of the standard
London model one wonders why the model which describes correctly the field of
vortices away of their cores, fails badly at low temperatures. From the point
of view of a consistent London description, the only suspicious point in
deriving the free energy (\ref{F_Lond}) and the corresponding magnetization
(\ref{M_Lond}) is the cutoff employed which generates the term $\ln(H_{c2}/B)$.

\subsection{London model modified to accommodate $\bm{\xi(B)}$}

Hence, we write the free energy in the form:
\begin{align}
{\tilde F}=\frac{\phi_{0}B } {32\pi^{2}\lambda^{2} }\, \ln\frac{ e \eta{\tilde
H}(B) }{ B }\,,\qquad{\tilde H}=\frac{\phi_{0}}{2\pi\xi^{2}(B)}\,, \label{F1}%
\end{align}
where $\xi(B)$ is the \textit{field dependent cutoff} (the core size).
Clearly, $\xi(H_{c2})$ is the standard coherence length associated with
$H_{c2}$, so that ${\tilde H} (H_{c2})=H_{c2}$. Then, evaluating
$M=-\partial{\tilde F}/\partial B$, we obtain:
\begin{align}
M=-\,\frac{\phi_{0} }{32\pi^{2}\lambda^{2} }\left(  \ln\frac{ \eta{\tilde H}
}{ B }\,+ \frac{B}{{\tilde H}}\,\frac{d{\tilde H}}{dB}\right)  \,. \label{M1}%
\end{align}

The idea of the following manipulation is to find a field ${\tilde H}$ that
generates the form (\ref{M_exp}), or in other words, that represents the
experimental data. After equating (\ref{M1}) and (\ref{M_exp}), one can solve
a linear differential equation for ${\tilde H}(B)$ with the boundary condition
${\tilde H}(H_{c2})=H_{c2}$. The result in terms of ${\tilde h}={\tilde
H}/H_{c2}$ and $b=B/H_{c2}$ reads:
\begin{equation}
\ln{\tilde h}=\frac{\alpha}{b}\,\ln b+\frac{(\ln\eta+\alpha)(1-b^{2})}{2b} \,.
\label{solution}%
\end{equation}
This corresponds to the normalized cutoff distance (the core radius)
\begin{align}
\xi^{*}(B) = \frac{\xi(B) }{ \xi_{c2} }= b^{-\alpha/2b}\, \exp\frac{(\ln
\eta+\alpha)( b^{2}-1)}{4b} . \label{xi(B)}%
\end{align}
It is readily shown that the slope of $\xi^{*}(b)$ at $H_{c2}$ is determined
by the parameter $\eta$:
\begin{align}
\frac{d\xi^{*}}{db}\Big|_{b=1} = \frac{1}{2}\, \ln\eta\,. \label{xi'(1)}%
\end{align}
Hence, when the field decreases from $H_{c2}$, $\xi(B)$ decreases for $\eta>1
$ and increases for $\eta<1$.

\begin{figure}
[ptb]
\begin{center}
\includegraphics[
height=7.563cm,
width=9.1028cm
]%
{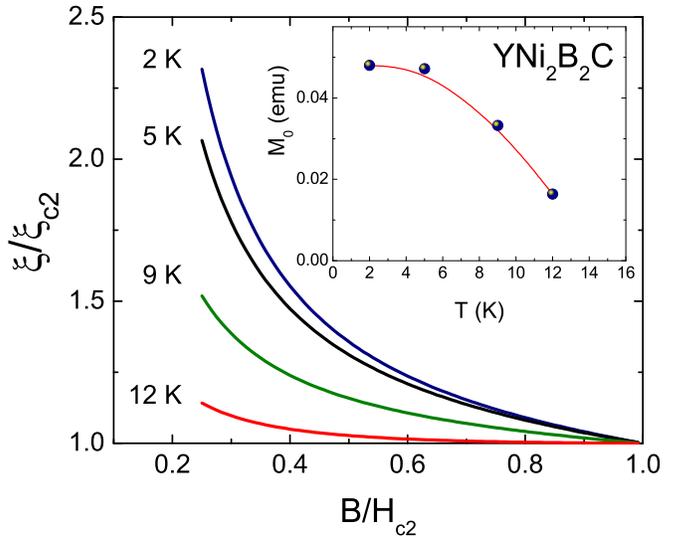}%
\caption{(color online) $\xi(B)$ extracted from the data of Fig.\thinspace
\ref{fYNBC} for YNi$_{2}$B$_{2}$C with the help of Eq.\thinspace(\ref{xi(B)}).
The upper curve is for $T=2\,$K, the lowest is for 12\thinspace K. The inset
shows $M_{0}(T)$ extracted from the fits of Fig.\thinspace\ref{fYNBC}. }%
\label{fYNBCxi}%
\end{center}
\end{figure}

Using Eq.\thinspace(\ref{xi(B)}), we can calculate the normalized cutoff
$\xi^{\ast}(b)$ responsible for deviations of $M(B,T)$ from the standard
London behavior for YNi$_{2}$B$_{2}$C shown in Fig.\thinspace\ref{fYNBC} since
we have $\eta$ and $\alpha$ representing these data sets (note that $M_{0}$
does not enter Eq.\thinspace(\ref{xi(B)})). The curves $\xi^{\ast}(b)$ for
$T=$ 2, 5, 9, and $12$\thinspace K calculated with Eq.\thinspace(\ref{xi(B)})
are shown in Fig.\thinspace\ref{fYNBCxi}.%

It is worth observing that $\xi(B)$ so obtained is qualitatively similar to
the $B$ dependence of the core size seen in $\mu$SR experiments.\cite{Sonier1}
Moreover, it is argued in Ref.\,\onlinecite{KZ2} that the field dependence of
$\xi$ should weaken with increasing temperature in accord with
Fig.\,\ref{fYNBCxi}.

Another point to make is only a moderate variation of $\xi$ which is needed to
account for strong deviations from the linear $M(\ln B)$. For example, at
2\,K, $\xi$ changes only by a factor of 2 over most of the mixed state field
domain. Since the cutoff enters the energy (\ref{F1}) under the log-sign, it
might be surprising that such a difference suffices to cause a drastic
deviation of the 2\,K curve in Fig.\,\ref{fYNBC} from a straight line. The
puzzle is resolved if one observes that the field dependence of $\xi$
translates to \textit{non-logarithmic} corrections to the standard London
magnetization, see Eq.\,(\ref{M1}).

\begin{figure}
[ptb]
\begin{center}
\includegraphics[
height=13.4368cm,
width=9.1028cm
]%
{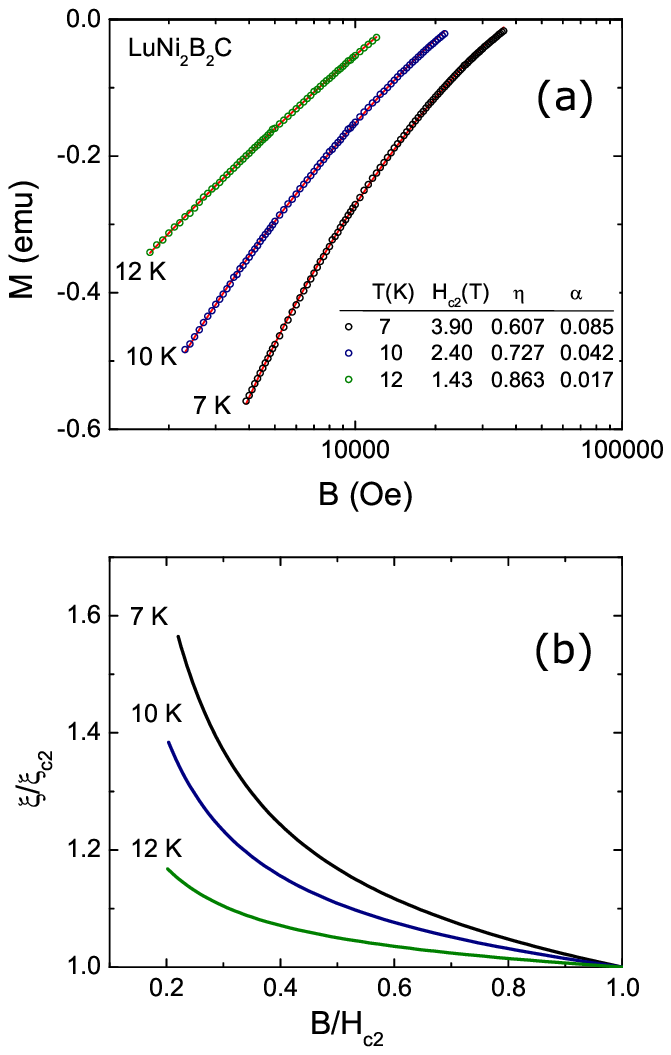}%
\caption{(color online) (a) $M(B)$ for LuNi$_{2}$B$_{2}$C for T=7,10 and 12 K.
(b) $\xi(B)$ corresponding to the graphs of the upper panel.}%
\label{LuT}%
\end{center}
\end{figure}

The same analysis has been applied to the magnetization data for a crystal of
LuNi$_{2}$B$_{2}$C [whose crystal structure and superconductive properties are
similar to YNi$_{2}$B$_{2}$C], yielding similar results as shown in
Fig.\,\ref{LuT}.

\subsection{ Lu(Ni$_{1-x}$Co$_{x})_{2}$B$_{2}$C}

As mentioned, Ref.\,\onlinecite{KZ2} argues that the field dependence of the
core size is weakened by increasing temperature and scattering. In order to
study the scattering dependence of $\xi(B)$ we turn to a series of crystals
Lu(Ni$_{1-x}$Co$_{x})_{2}$B$_{2}$C in which the mean-free path is
progressively reduced by increasing the Co content.\cite{Cheon}%

\begin{figure}
[ptbh]
\begin{center}
\includegraphics[
height=13.4917cm,
width=9.1028cm
]%
{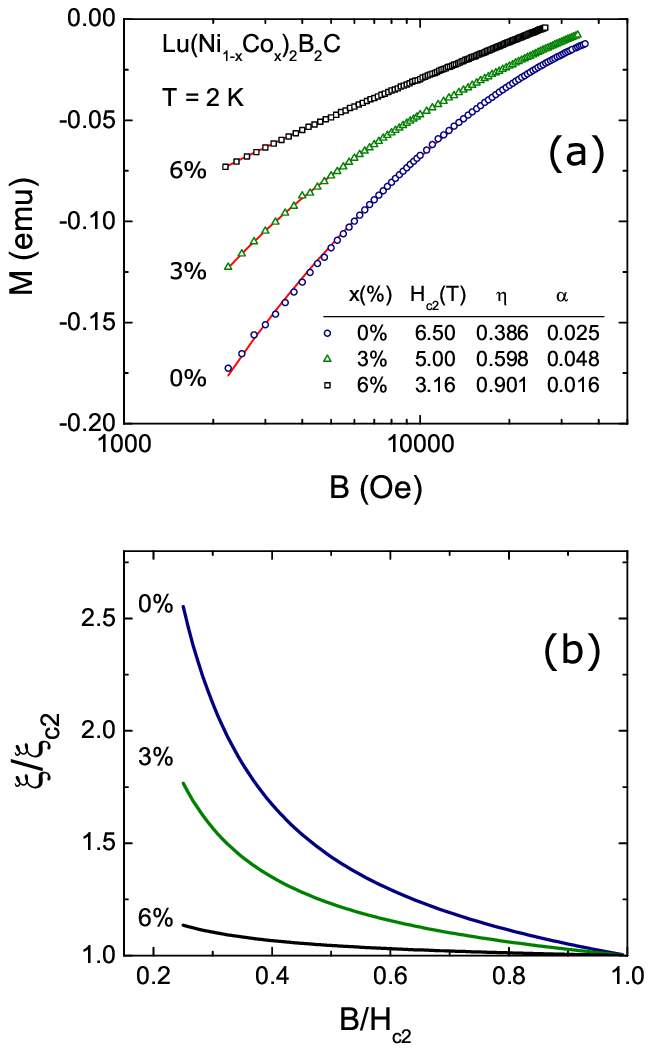}%
\caption{(color online) (a) $M(B)$ for Lu(Ni$_{1-x}$Co$_{x})_{2}$B$_{2}$C with
$x=0$, 3, and 6\%. (b) $\xi(B)$ corresponding to graphs of the upper panel.}%
\label{LuX}%
\end{center}
\end{figure}

In the upper panel of Fig.\thinspace\ref{LuX}, $M(B)$ is shown for $x=0$, 3,
and 6\% (for which $H_{c2}$ has been independently measured) at the same
temperature of 2\,K. The fit parameters $\eta$ and $\alpha$ are also shown and
the calculated field dependent core sizes are given in the lower panel.%

We note that for 6\% Co, the ratio of the zero-$T$ coherence length to the
mean-free path has been estimated in Ref.\,\onlinecite{Cheon} as exceeding 10,
which places this sample close to the dirty limit. The core size for this
crystal is seen to vary only by about 10\%, which is in accord with the
theoretical finding that field dependence of $\xi$ disappears in the dirty
limit.\cite{K85,KZ2}

\subsection{ NbSe$_{2}$}

\begin{figure}
[ptb]
\begin{center}
\includegraphics[
height=13.3533cm,
width=9.1028cm
]%
{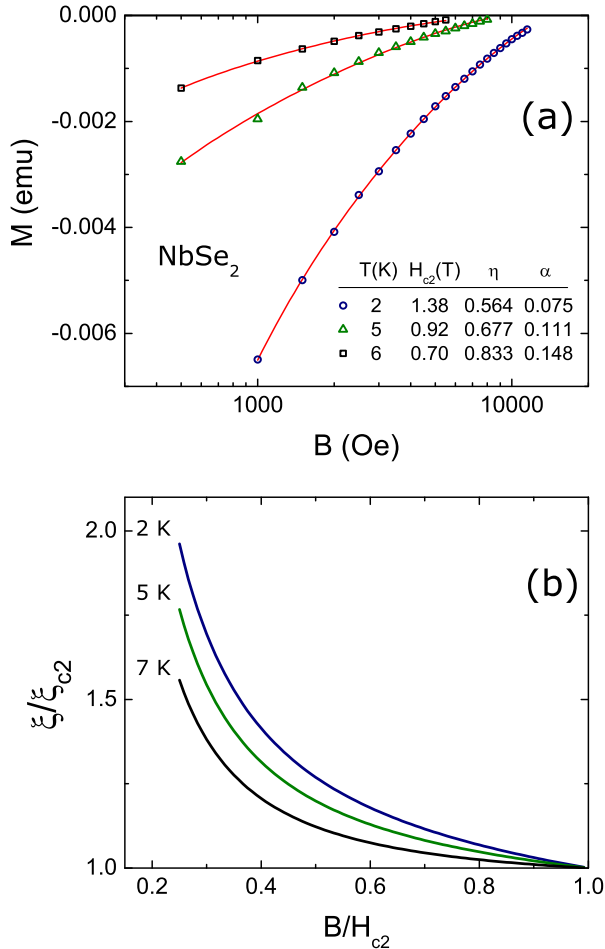}%
\caption{(color online) (a) $M(B)$ for NbSe$_{2}$. (b) $\xi(B)$
corresponding to graphs of
the upper panel.}%
\label{fNbSe2}%
\end{center}
\end{figure}

The superconducting anisotropy of this material is stronger than in
borocarbides discussed above. As is seen in Fig.\thinspace\ref{fNbSe2},
deviations of $M(\ln B)$ from the standard London linearity are profound along
with the corresponding field dependence of $\xi$.%

\subsection{ MgB$_{2}$}%

\begin{figure}
[ptbh]
\begin{center}
\includegraphics[
height=13.3159cm,
width=9.1028cm
]%
{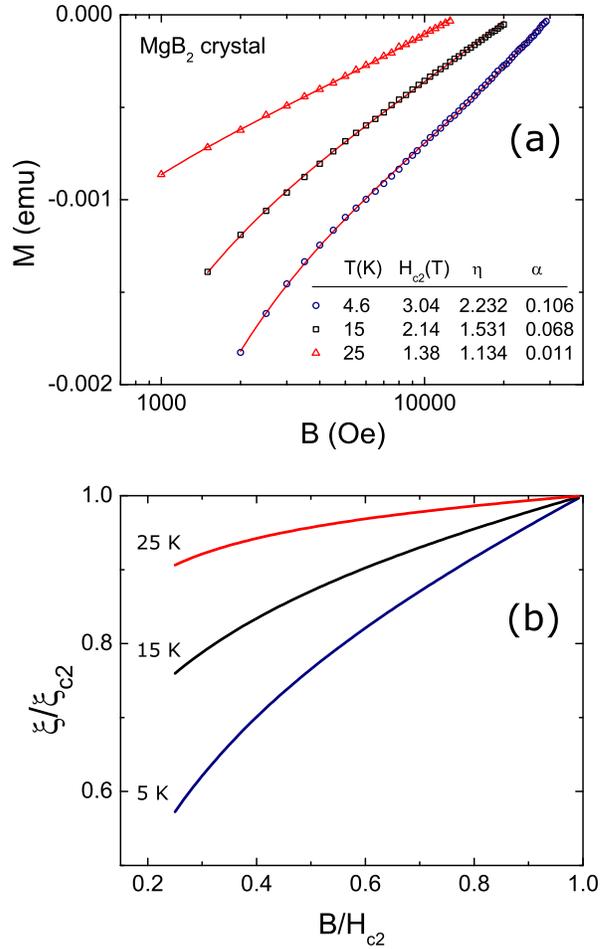}%
\caption{(color online) (a) MgB$_{2}$, $T=4.6,15$ and 25\thinspace K.(b)
corresponding $\xi(B)$.}%
\label{fMgB2}%
\end{center}
\end{figure}

The upper panel of Fig.~\ref{fMgB2} shows $M(\ln B)$ in fields parallel to the
$c$ axis of a single crystal MgB$_{2}$. One readily sees a qualitative
difference from the preceding examples: the curvature of $M(\ln B)$ for
$T=4.6$\thinspace K being negative in low fields becomes positive in large
fields. Still, we can fit well the data for all $T$'s to the form of
Eq.\thinspace(\ref{M_exp}) with parameters $\eta$ and $\alpha$ given in the table.

The most interesting feature is that the obtained values of $\eta$ exceed
unity. According to Eq.\,(\ref{xi'(1)}) this means that starting from $H_{c2}%
$, $\xi$ decreases with decreasing field. This is shown in the lower panel of
Fig.\,\ref{fMgB2}. We attribute this unusual behavior to the two-gap nature of
this material: the small gap on the $\pi$-sheet of the Fermi surface opens up
in decreasing fields thus causing a decrease of $\xi$.

\begin{figure}
[ptbh]
\begin{center}
\includegraphics[
height=13.2347cm,
width=9.1006cm
]%
{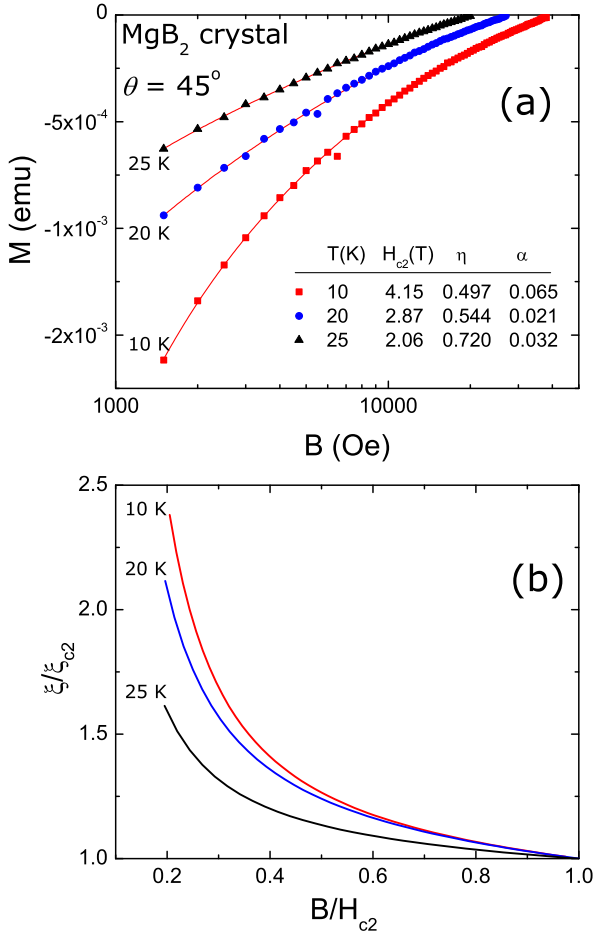}%
\caption{(color online) (a) Magnetization of the MgB$_{2}$ crystal in applied
field at $45^{\circ}$ to the $c-$ axis. (b) corresponding coherence length.
Note the difference in the behavior of $\xi^{\ast}(b)$ from the case of the
field along $c$ of Fig.\thinspace\ref{fMgB2}.}%
\label{45degr}%
\end{center}
\end{figure}

If indeed the unusual behavior of $\xi(B)$ for MgB$_{2}$ is due to suppression
of the small gap in fields of few kG for $H||c$, and if the suppression field
is \textit{isotropic}, then by going to other field orientation away of the
$c$ axis, we can push the effect of the small gap out of the high field domain
of our interest. To check this hypothesis we acquired the data for the applied
field at $45^{\circ}$ to the $c$ axis where $H_{c2}(0)$ is accessible with our
equipment. Figure \ref{45degr} shows the result similar to that for Y and
Lu-based borocarbides. This suggests that, e.g., for $T=10$\thinspace K, in
the field domain examined (from $H_{c2}=4.15$\thinspace T down to about
0.4\thinspace T or $b\approx0.1$) the small gap is not yet fully formed.%

Of course, this interpretation is much too simple because for other than
$H\,||\, c$ orientation the strong anisotropy of $\xi$ should be taken into
account, the subject of our future work It should also be noted here that the
macroscopic phenomenology of magnetic properties of MgB$_{2}$ is still
debated. Despite the two-band nature of this material, within the London
approach, we employ \textit{one} penetration depth $\lambda$ and \textit{one}
cutoff length $\xi$ when describing the vortex lattice in reciprocal space.
Judging by literature, this point of view is not universally shared by all
MgB$_{2}$ community.

\subsection{ V$_{3}$Si}

Given the slope $dH_{c2}/dT=19.4$\thinspace T/K and $T_{c}=16.6$\thinspace K,
it is likely that the low temperature upper critical field of this material
exceeds 20\thinspace T; no direct measurements of $H_{c2}$ were conducted in
this field range. One might treat $H_{c2}$ as an extra fitting parameter to be
extracted from the data on $M(B)$. However, the numerical procedure of
extracting both $H_{c2}$ and $\eta$ from the magnetization data is unstable
because their product enters the formulas we use. For this reason we consider
here only $M(B)$ for $T>13$\thinspace K for which $H_{c2}$ was measured.%

\begin{figure}
[ptb]
\begin{center}
\includegraphics[
height=6.6887cm,
width=9.1028cm
]%
{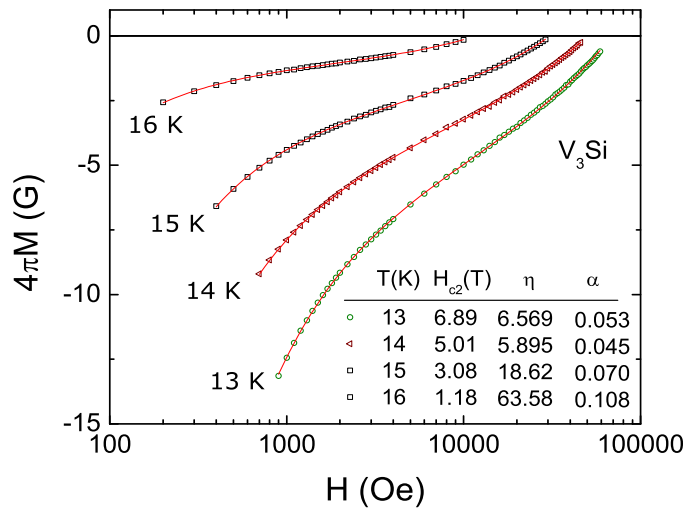}%
\caption{(color online) Magnetization of the V$_{3}$Si single crystal at
$T=13,14,15$ and 16\thinspace K.}%
\label{fV3Si}%
\end{center}
\end{figure}

It is worth noting that Eq.~(\ref{M_exp}) is good enough even for a quite
unusual shape of $M(\ln B)$ in this material: the curvature of $M(\ln B)$
changes sign in all data we have examined. With the help of Eq.\,(\ref{M_exp})
we readily find that the inflection point is at $b_{i}=\sqrt{\alpha
/(\alpha+\ln\eta)}$. As is seen from the table of Fig.\,\ref{fV3Si}, for all
curves given, $\alpha\ll\ln\eta$ which leads to
\begin{equation}
\ln b_{i} \approx\frac{1}{2}\left(  1+\frac{\alpha}{\ln\eta}\right)  \,.
\label{inflection}%
\end{equation}
Hence the inflection points of $M(\ln b)$ are approximately in the same place,
$\ln b_{i}\approx0.5$, for all temperatures.\cite{remark1} Therefore, the
curves $M(\ln B)$ are concave for $b<b_{i}\approx0.16$, i.e., at fields hardly
in the domain of applicability of our high-field model. On the other side of
the inflection point where the curves $M(\ln B)$ are convex, we may deal with
the Abrikosov domain where $M$ is linear in $H_{c2}-B$; it is easy to see then
that $M(\ln B)$ should be convex.\cite{remark2} Application of the London
approach in this domain cannot be justified. Therefore, there is no point in
trying to extract $\xi(B)$ from the data on V$_{3}$Si (formally, since all
$\eta$'s in the table of Fig.\,\ref{fV3Si} exceed unity, this extraction would
have given $\xi(B)$ decreasing with decreasing field which would have
contradicted the $\mu$SR data of Ref.\,\onlinecite{Sonier_V3Si}).\newline

\section{Discussion}

The main point of this work is to argue that the field dependence of the core
size is a generic low-temperature property of all sufficiently clean
superconductors. Moreover, incorporating this dependence in the London
approach broadens considerably its applicability for describing macroscopic
reversible magnetic properties. Still, the empirical approach adopted here
lacks microscopic justification. If proven correct, the modified London scheme
calls for revisiting many results obtained within the standard London model in
which the field independent core size or the cutoff are involved. Field
dependencies of the flux-flow resistivity, of the specific heat in the mixed
state, or the core pinning are some examples.

\begin{figure}
[ptbh]
\begin{center}
\includegraphics[
height=7.3104cm,
width=9.1028cm
]%
{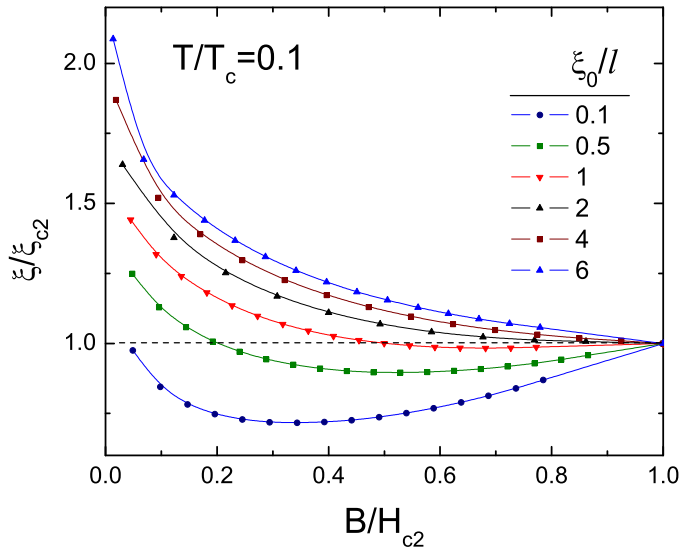}%
\caption{(color online) The field dependence of $\xi$ calculated
microscopically for $T/T_{c}=0.1$ and for a few scattering parameters $\xi
_{0}/\ell$ where $\xi_{0}$ is zero-$T$ BCS coherence length and $\ell$ is the
mean-free path for scattering on non-magnetic impurities.}%
\label{fmiranovich}%
\end{center}
\end{figure}

One can foresee a number of difficulties in trying to develop such a
justification. The cutoff size we extract from $M(B)$ data is not necessarily
the same as the core size defined as being proportional to the slope of the
order parameter at the vortex axis: approaching the core from the outside to
determine the cutoff we may have a different result than when examining the
core structure starting from the core center. Having this in mind, it is not
surprising that the microscopic calculations of $\rho_{c}\propto
(d\Delta/dr|_{r\rightarrow0})^{-1}$ in Ref.\thinspace\onlinecite{Pedja} do not
agree in detail with our empirical results (to our knowledge, this is the only
calculation of this sort that has been tried to date). Fig.\thinspace
\ref{fmiranovich} shows the output of this calculation normalized in the
manner of this paper for a model isotropic material at $T=0.1\,T_{c}$ and for
a few values of the scattering parameter $\xi_{0}/\ell$ ($\xi_{0}=\hbar
v_{F}/\pi\Delta(0)$ is the zero-$T$ BCS coherence length and $\ell$ is the
mean-free path for the non-magnetic scattering). We note that while the curves
generated are mostly qualitatively similar to what we extract from the
magnetization data, this calculation does not confirm our assertion about
weakening of the field dependence of the core size with increasing scattering.%

A different approach to evaluation of the core size is chosen in
Ref.\,\onlinecite{KZ2}: it is argued on physical grounds that since $\Delta
\to0$ at the vortex center and the field is practically uniform within the
core for $\kappa\gg1$, one can use the Helfand-Werthamer\cite{HW}
linearization technique employed for calculation of $H_{c2}$ also for the core
size in the high-field mixed state. Within this approach, the order parameter
near the vortex center satisfies a linear equation $-\xi^{2} \Pi^{2}%
\Delta=\Delta$, where $\bm
\Pi=\bm\nabla+2\pi\bm A/\phi_{0}$ and $\xi$ is found solving the basic BCS
self-consistency equation. This produces $\xi(T,\ell;B)$ in qualitative
agreement with what we extract from the magnetization data in this work (for
all cases other than V$_{3}$Si and MgB$_{2}$ in field along the $c$ axis); in
particular, the analytical field dependence of $\xi$ obtained in this way
disappears when $T\to T_{c}$ or $\ell/\xi_{0}\to0$.

In our view, the question still remains which of these two theoretical
approaches, Ref.\,\onlinecite{Pedja} or Ref.\,\onlinecite{KZ2}, describes
better various data on $\xi(B)$. An important role in resolving the question
belongs to experimental studies of how the field dependence of the core size
affects other physical properties of the mixed state.

\subsection{Flux-flow resistivity}

As an example of such a cross-examination we took data on the flux-flow
resistivity $\rho_{f}$ from measurements of the microwave surface impedance of
YNi$_{2}$B$_{2}$C.\cite{Marcin} The data show large deviations of the measured
$\rho_{f}$ from the Bardeen-Stephen linear field dependence, $\rho_{f}%
=\rho_{n}B/H_{c2}$. This formula is obtained assuming a field independent core
size $\xi=\xi_{c2}=\sqrt{\phi_{0}/2\pi H_{c2}}$. Clearly, if $\xi$ does depend
on the field, one has:
\begin{equation}
\frac{\rho_{f}}{\rho_{n}}=B\,\frac{2\pi\xi^{2}(B)}{\phi_{0}}=\frac{B}{H_{c2}%
}\,\frac{\xi^{2}(B)}{\xi_{c2}^{2}}=b\,\xi^{\ast2}(b)\,. \label{rhof}%
\end{equation}
Hence, for each data set $\rho_{f}(B)$, we can extract
\begin{equation}
\frac{\xi(B)}{\xi_{c2}}=\sqrt{\frac{\rho_{f}(B)}{\rho_{n}\,b}}\,;
\label{xi-rho}%
\end{equation}
in other words, we can delegate deviations from $\rho_{f}\propto B$ to the
field dependence of $\xi$ and see whether or not the obtained $\xi(B)$ agrees
with that extracted from magnetization data.

\begin{figure}
[ptbh]
\begin{center}
\includegraphics[
height=7.1302cm,
width=9.1028cm
]%
{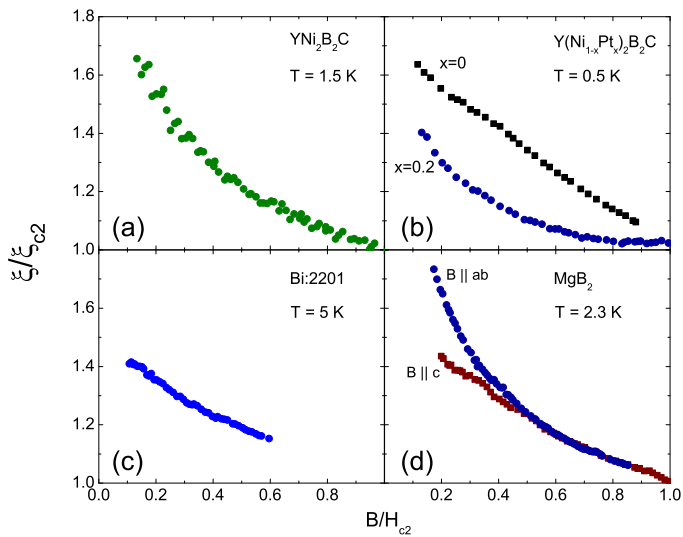}%
\caption{(color online) $\xi(b)/\xi_{c2}$ versus $b=B/H_{c2}$ for Y(Ni$_{1-x}%
$Pt$_{x})_{2}$B$_{2}$C with $x=0$ and 0.2 extracted from the flux-flow
resistivity data of Ref.\thinspace\onlinecite{flow} as explained in the text.}%
\label{fflow}%
\end{center}
\end{figure}

Utilising the flux-flow resistivity data of Fig.\,3 of
Ref.\,\onlinecite{Marcin} and applying Eq.\,(\ref{xi-rho}), we obtain the
result shown in the panel (a) of Fig.\,\ref{fflow}. Comparing it with our
Fig.\,\ref{fYNBCxi} for the same material, we obtain reasonable agreement,
notwithstanding the usage of different samples in these two experiments.

As another example, two selections of data for Y(Ni$_{1-x}$Pt$_{x})_{2}$%
B$_{2}$C from Fig.\thinspace3 of Ref.\thinspace\onlinecite{flow} were used to
calculate $\xi(B)/\xi_{c2}$, as shown Fig.\thinspace\ref{fflow}(b). We see
that in addition to the expected decrease with field, $\xi(B)$ is suppressed
by increasing scattering (i.e., increasing impurity content of Pt) again in
accord with the examples discussed above.

Panel (c) of Fig.\,\ref{fflow} shows the result of the same exercise with a
d-wave material (an overdoped crystal of Bi-2201).\cite{Matsuda} This example
supports the idea that the field dependence of the London cutoff is a generic
feature of type-II superconductors with no direct relation to the order
parameter symmetry.

The last panel of Fig.\,\ref{fflow} presents the cutoff $\xi(b)/\xi_{c2}$ for
two field orientations of MgB$_{2}$ extracted from the flux-flow resistivity
data of Ref.\,\onlinecite{Matsuda1}, again using Eq.\,(\ref{xi-rho}). For the
field along $ab$, both $M(B)$ and $\rho_{f}(B)$ data yield qualitatively
similar results. This, however, is not the case for the field along $c$ as
evident by comparing Figs.\,\ref{fMgB2} and \ref{fflow}(d). Thus it appears
that for the two-gap MgB$_{2}$ with its greater complexity, the simple scheme
of incorporating a field-dependent cutoff to the London model does not work
for all field orientations.%

\subsection{On nonlocality}

Deviations in $M(\ln B)$ from the standard London linear behavior have been
thought to come from the effects of the nonlocal relation between current and
the vector potential inherent for superconductors.\cite{nonloc} The nonlocal
corrections to London theory turned out to be an effective tool in describing
evolution and transitions in vortex lattice structures.\cite{review} However,
it is difficult at this stage to sort out what part of the deviations of $M$
from linear-in-$\ln B$ behavior arises solely from the core-size field
dependence and what part should be relegated to the nonlocality. In
particular, the difficulty comes from analogous weakening of the two effects
with increasing temperature and scattering strength. Looking back to the
overall satisfactory data analysis of $M(\ln B)$ for Lu(Ni$_{1-x}$Co$_{x}%
)_{2}$ B$_{2}$C of Ref.\,\onlinecite{Kogan-Budko} based on nonlocal effects,
we note that the analysis produced an excessively rapid reduction of the
"nonlocality range" with temperature for samples with elevated impurity
content. We have to conclude that more high precision experimental results are
needed before a conclusive judgement is made on the relative importance of
contributions to $M(B)$ due to the nonlocality and to the field dependence of
the core size.

\subsection{Why $\lambda(B)$ should be used with caution within London theory}

As mentioned above, the basic London equation (\ref{London}) along with the
energy expression (\ref{F_Lond}) for intermediate fields (or Eq.\,(\ref{F1})
with an unspecified cutoff) imply that the London penetration depth $\lambda$
is a temperature dependent \textit{constant} in a homogeneous material.
Assuming a field dependent $\lambda$ would have changed the London equation
(\ref{London}) \textit{per se}: the quantity $\lambda^{2}(h)$ cannot be taken
out of differentiation operators. As a result, the Fourier components of the
solution $h(\bm k)$ for a single vortex would have been different from
$\phi_{0}/(1+\lambda^{2}k^{2})$ and the energy (\ref{F_Lond}) would have been
different as well. Therefore, unlike the case of the cutoff $\xi$, relaxing
the requirement of a constant $\lambda$ causes basic changes in the London
approach and therefore should not be taken lightly. It also worth recalling
that a constant $\lambda$ is derived from the microscopic BSC theory (as
$\bm k\to0 $ limit of the BCS kernel in the nonlocal connection between the
persistent current and the vector potential). To our knowledge, there is no
microscopic justification for a field-dependent $\lambda$ (in non-magnetic
superconductors). In other words, the London theory is rigid with respect to a
constancy of $\lambda$, unlike the case of $\xi$.

Yet, quite often, analyzing data with the help of the London model (i.e.,
starting with a constant $\lambda$) it is concluded that $\lambda$ should be
field dependent. Numerous examples are found in the literature on $\mu$SR
(see, e.g., Ref.\thinspace\onlinecite{Sonier1}) and in many recent
publications on MgB$_{2}$.\cite{Angst,Cubitt,Eist,Klein} An \textquotedblleft
operational" justification implied for this apparent contradiction usually
goes like this: of course, the field distribution for a single vortex is
described by Eq.\thinspace(\ref{London}) with a constant $\lambda$. However,
in the mixed state the average order parameter is suppressed by overlapping
vortex fields, and therefore, the \textquotedblleft macroscopic" $\Lambda$
($\neq$ to $\lambda$ calculated for $H\rightarrow0$) should enter the free
energy Eq.\thinspace(\ref{F_Lond}). This macroscopic parameter may depend on
the average magnetic field $B$.

The inconsistency of such an argument is exposed if one considers the clean
limit at zero temperature. In this case the London $\lambda$ does not depend
on the order parameter (in fact, it depends only on the total electron
density), so that the mixed-state order parameter suppression cannot be
reffered to as a general justification for employing $\Lambda(B)$.

\subsection{On the superfluid density}

The quantity $\lambda^{2}(0)/\lambda^{2}(T)$ is often taken as a measure of
the superfluid density $n_{s}$. This assignment has unambiguous justification
only for isotropic superconductors (see, e.g., Ref.\,\onlinecite{Abrik}) and
when $\lambda(T)$ is defined as the penetration depth of a small magnetic
field (strictly speaking in the limit $H\to0$). One of the attractive features
of the standard London theory is that one can extract $1/\lambda^{2}(T)$
directly from the magnetization (\ref{M_Lond}) by measuring the constant slope
$dM/d(\ln B)$ for each temperature. This way of determining the superfluid
density $n_{s}$ rests, therefore, upon whether or not the standard London
model for $M$ is valid. As we have seen in a number of examples above, this is
quite often not the case.

Perhaps the best example of the futility of extracting $n_{s}$ from
magnetization data is provided by the data for V$_{3}$Si. As is seen in
Fig.\,\ref{fV3Si}, e.g., for $T=14\,$K, the slope $dM/d(\ln B)$ decreases with
the field increasing up to $\sim1\,$T, but it grows with further field
increase toward $H_{c2}$. If one takes literally the proportionality between
$n_{s}$ and the slope $dM/d(\ln B)$, one should conclude that the field
suppresses the ``superfluid density" as long as $B$ is under $\approx1\,$T but
for $B>1\,$T, $n_{s}$ is enhanced by increasing field.

Hence, extracting any quantitative information about the superfluid density or
the penetration depth from the magnetization data with the help of the
\textit{standard} London model should be taken ``with a grain of salt" at best
even for such ``simple" materials as V$_{3}$Si, not to speak about MgB$_{2}$
in which the different field behavior of the two gaps further complicates the
matter.\cite{Angst,Eist,Klein}

To summarize, the field dependence of the core size $\xi(B)$ has been
incorporated in the London model to describe intermediate field reversible
magnetization $M(B,T)$ for materials with large $\kappa$. The dependence
$\xi(B)$ is directly related to deviations in $M(\ln B)$ from linear behavior
prescribed by the standard London model. A method to extract $\xi(B)$ from the
magnetization data is proposed. For most materials examined, $\xi(B)$ so
obtained decreases with increasing field; the dependence becomes weaker with
increasing temperature or with strengthening the non-magnetic scattering, in
qualitative agreement with theoretical predictions and with existing $\mu$SR,
SANS, and flux-flow resistivity data. The method, however, fails when applied
to MgB$_{2}$ and - surprisingly - to V$_{3}$Si, the subject for a separate discussion.

\section{Acknowledgements}

The authors thank Morten Eskildsen for useful discussions and for sharing SANS
data on CeCoIn$_{5}$ before publication. Ames Laboratory is operated for US
DOE by the Iowa State University under Contract No. W-7405-Eng-82. Research at
ORNL was sponsored by the Division of Materials Sciences and Engineering,
Office of Basic Energy Sciences, U.S. Department of Energy, under contract
DE-AC05-00OR22725 with Oak Ridge National Laboratory, managed and operated by
UT-Battelle, LLC. R.P. acknowledges a support from NSF Grant number
DMR-05-53285 and from the Alfred P. Sloan Foundation.


\begin{thebibliography}{99}
\bibitem {remark*}Eq.\,(\ref{M_exp}) is in fact a general expansion of a
function singular as $b=B/H_{c2}\to0$: $c_{1}/b+c_{2}\ln b+c_{3}+c_{4}b+...$
\bibitem {remark1}This suggest a possibility of a certain scaling procedure as
a result of which all $M(\ln b)$ of Fig.\,\ref{fV3Si} nearly collapse to a
single curve. We will not dwell on this matter, since the physical meaning of
such a scaling is still unclear.
\bibitem {remark2}One can calculate the linear slope $dM/dB|_{B\to H_{c2} }$
using Eq.\,(\ref{M_exp}) to obtain $(1+2\alpha+\ln\eta)/16\pi\kappa^{2}$. The
Abrikosov's slope is $\approx1/8\pi\kappa^{2}\beta_{A}$ with $\beta_{A}=1.16$
for a triangular lattice at $T\to T_{c}$. It is now easy to see that
Eq.\,(\ref{M_exp}) with parameters $\eta$ and $\alpha$ from the table in
Fig.\,\ref{fV3Si} for $T=16$\,K (which is close to $T_{c}\approx16.8$\,K)
gives the slope of the same order as that calculated according to Arikosov.
This suggests that in V$_{3}$Si the Abrikosov domain might be very broad. We
note that if $\rho_{c}$ would behave as $\approx\xi_{c2}/\sqrt{b}$ in a broad
domain under $H_{c2}$, the intervortex distance would have been on the order
of $\rho_{c}$; in other words, one would expect Abrikosov's linear $M(B)$ in
this domain. This speculation is worth of experimental examination.
\end{thebibliography}
\end{document}